\documentstyle[preprint,aps]{revtex}
\begin{document}
\tighten
\draft
\preprint{PSU/TH/175; hep-ph/9610443}
\title {ONE MORE ANALYTIC RESULT FOR  $\alpha^{2}(Z\alpha)^{5}m$ CORRECTION 
TO THE LAMB SHIFT} 
\author {Michael I. Eides \thanks{E-mail address:  
eides@phys.psu.edu, eides@lnpi.spb.su}}
\address{ Department of Physics, Pennsylvania 
State University, 
University Park, PA 16802, USA\thanks{Temporary address.}\\ 
and
Petersburg Nuclear Physics Institute,
Gatchina, St.Petersburg 188350, Russia\thanks{Permanent address.}}
\author{Howard Grotch\thanks{E-mail address: h1g@psuvm.psu.edu}}
\address{Department of Physics, Pennsylvania State University,
University Park, PA 16802, USA}
\author{Valery A. Shelyuto \thanks{E-mail address: 
shelyuto@onti.vniim.spb.su}} 
\address{D. I.  Mendeleev Institute of Metrology, 
St.Petersburg 198005, Russia}
\date{October, 1996}

\maketitle
\begin{abstract}
The analytic result for the radiative correction of order 
$\alpha^{2}(Z\alpha)^{5}m$ to the Lamb shift connected with a polarization 
insertion in one of the two external Coulomb lines is obtained. This 
correction arises from a gauge invariant set of diagrams which contain, 
besides the polarization insertion in the Coulomb leg, all one-loop 
radiative photon insertions in the electron line with two external Coulomb 
lines.  
\end{abstract} 

Theoretical work on high order corrections to hyperfine splitting (HFS)
and Lamb shift in hydrogenlike ions was concentrated recently on calculation 
of all nonrecoil contributions of order $\alpha^2(Z\alpha)^5$.  This work is 
now successfully completed, and all calculations for the hyperfine splitting 
\cite{eks1,eks2,eks5,eksl,kn,es} and for the Lamb shift 
\cite{ego,eg,eg4,pach1,eg5,eksl,pach2,es} were performed 
independently by two different groups, and the results of these calculations 
are in excellent agreement.

Still there exists a certain asymmetry between the results for HFS and 
Lamb shift with respect to our knowledge of analytical results. For both HFS 
and Lamb shift contributions of order $\alpha^2(Z\alpha)^5$ are induced by 
six gauge invariant sets of diagrams in Fig.1, which correspond to different 
radiative corrections to the skeleton graph with two Coulomb photons 
attached to the electron line \cite{eks1,ego}. While corrections to HFS 
induced by the first three sets of diagrams in Figs.1a-1c are known in 
analytic form \cite{eks1,eks2}, in the case of the Lamb shift analytic 
results exist only for the first two sets of diagrams in Fig.1 \cite{lap}. 
It is the aim of the present paper to obtain the analytic result for the 
contribution to the Lamb shift induced by the gauge invariant set of 
diagrams in Fig.1c, containing insertions of one radiative photon in the 
electron line and simultaneous insertion of a one-loop polarization operator 
in one of the Coulomb lines. This would imply that contributions of the 
diagrams in Figs.1a-1c are then known analytically for both HFS and Lamb 
shift.

As  was discussed at length in Ref.\cite{ego} the respective
contribution to the Lamb shift of $S$ states is given  by the expression

\begin{equation}            \label{generalform}
\Delta E= -\frac{\alpha^2\left(Z\alpha\right)^5}{\pi n^3} 
\:{m}\left(\frac{m_r}{m}\right)^3 \frac{32}{\pi^2} \int_0^\infty 
dk\:{I}(k)\:L(k)\:,  
\end{equation}

where

\begin{equation}
{I(k)}= \int_0^1 dv \frac{v^2(1-v^2/3)}{4+(1-v^2)k^2}\:
\end{equation}  

describes the Coulomb line with one polarization insertion, $L(k)$
corresponds to the sum of radiative corrections to  the electron line in
Fig.1, and $k=|{\bf k}|$ is the magnitude of the spatial momentum of the
external Coulomb photons measured in units of electron mass\footnote {All
integration momenta below are measured in units of electron mass.}.   

Explicit expressions for the electron factor $L(k)$ with on-mass-shell 
external electron lines in Eq.(\ref{generalform}) were obtained earlier in 
\cite{bg,eg,eg4} in the form of apparently different two-dimensional 
integrals over the Feynman parameters. Of course, one may prove that all 
these expressions coincide. 

For our present goals we need an expression for the 
electron factor, not in the form of the parametric integral, but as an 
explicit function of the exchanged momentum. We have started with the 
parametric expression in \cite{eg4}, and performed all parametric 
integrations explicitly. The initial expression for the electron factor in 
\cite{eg4} contains an auxiliary infrared regularization parameter 
$\lambda$, which was needed to make separate parametric integrals well 
defined. It is well known that the total expression for the electron factor 
is infrared finite and should not depend on the infrared photon mass 
$\lambda$. In accordance with this general statement the result of our 
parametric integration is nonsingular in $\lambda$ and admits the limit of 
vanishing $\lambda$. Explicitly, after tedious calculations we obtain 

\begin{equation}   \label{elfact}
L(k)=-\frac{1}{4}+\frac{1}{2}\ln{k^2}+\frac{1}{8}\frac{k^2}{1-k^2}\ln{k^2}
-\frac{\sqrt{k^2+4}}{2k}\ln\frac{\sqrt{k^2+4}+k}{\sqrt{k^2+4}-k}
\end{equation}
\[
+\frac{1}{k\sqrt{k^2+4}}\ln\frac{\sqrt{k^2+4}+k}{\sqrt{k^2+4}-k}
-3\left[\frac{1}{k^2}-\frac{\sqrt{k^2+4}}{2k^3}
\ln\frac{\sqrt{k^2+4}+k}{\sqrt{k^2+4}-k}\right]
\]
\[
+\frac{k}{8}\Phi(k)+\frac{1}{2k}\Phi(k)-\frac{2}{k^2}
\left[\frac{1}{k}\Phi(k)+\ln k^2-1\right]
\equiv \sum_{i=1}^{i=9}L_i(k),
\]

where 

\begin{equation}
\Phi(k)=k\int_0^1\frac{dz}{1-k^2z^2}\ln\frac{1+k^2z(1-z)}{k^2z}.
\end{equation}

This last integral in the definition of function $\Phi(k)$ may be also 
calculated in closed form in terms of dilogarithms, but the integral 
representation is more convenient for further calculations.

The expression for the spin-independent electron factor relevant for the 
Lamb shift calculation in Eq.(\ref{elfact}), has the same general structure 
as the respective expression for the spin-flip electron factor relevant for 
the calculation of the contribution to HFS \cite{eks1}. The same auxiliary 
function $\Phi(k)$ emerged also in the case of HFS.  

It is easy to obtain high- and low-frequency asymptotes of the electron 
factor   in Eq.(\ref{elfact})

\begin{equation}
L(k)_{k\rightarrow 0}=\frac{2}{3}\ln k^2-\frac{5}{9},
\end{equation}
\[
L(k)_{k\rightarrow\infty}=-\frac{1}{k^2}(\frac{1}{3}\ln k^2+\frac{35}{36}).
\]

For analytic calculation we also need the explicit expression for the 
polarization operator as the function of the exchanged momentum

\begin{equation}
{I(k)}=\frac{\sqrt{k^2+4}}{k^3}\left[\frac{1}{2}\ln\frac{\sqrt{k^2+4}+k}
{\sqrt{k^2+4}-k}-\frac{k}{\sqrt{k^2+4}}\right]
\end{equation}
\[
-\frac{1}{3}\frac{(k^2+4)^\frac{3}{2}}{k^5}
\left[\frac{1}{2}\ln\frac{\sqrt{k^2+4}+k}
{\sqrt{k^2+4}-k}-\frac{k}{\sqrt{k^2+4}}
-\frac{1}{3}\frac{k^3}{(k^2+4)^\frac{3}{2}}\right]
\equiv I_{11}(k)+I_{12}(k).
\]

Then the contribution to the Lamb shift induced by the diagrams in Fig.1c 
may be written in the form

\begin{equation} \label{repr}
\Delta E=-\frac{\alpha^2\left(Z\alpha\right)^5}{\pi n^3} 
\:{m}\left(\frac{m_r}{m}\right)^3 \frac{32}{\pi^2} \int_0^\infty 
dk\:\sum_{i=1}^{i=9}\sum_{j=1}^{j=2}{I}(k)\:L_i(k){I}_{1j}(k)\:
\end{equation}
\[
\equiv \frac{\alpha^2\left(Z\alpha\right)^5}{\pi n^3} 
\:{m}\left(\frac{m_r}{m}\right)^3 {32}
\:\sum_{i=1}^{i=9}\sum_{j=1}^{j=2}\delta\epsilon_{ij}.
\]

Calculation of the separate integrals for the terms $\delta\epsilon_{ij}$ is 
performed directly. We always avoid direct use of the function $\Phi(k)$ 
with the help of integrating by parts, exploiting the fact that the 
derivative of $\Phi(k)$ may be easily calculated in terms of elementary 
functions

\begin{equation}
\Phi'(k)=-\frac{\ln k^2}{1-k^2}-\frac{2}{k\sqrt{k^2+4}}
\ln\frac{\sqrt{k^2+4}+k}{\sqrt{k^2+4}-k}-\frac{k}{\sqrt{k^2+4}}.
\end{equation}

Explicit results for the separate terms on the right hand side in 
Eq.(\ref{repr}) are collected in the Table. In the process of 
calculation two auxiliary integrals

\begin{equation} 
\int_0^\infty\frac{kdk}{(1-k^2)\sqrt{k^2+4}}\ln 
k\cdot\ln\frac{\sqrt{k^2+4}+k}{2} 
=-\frac{\pi^2}{\sqrt5}\ln\frac{1+\sqrt5}{2}, 
\end{equation} 
\[ 
\int_0^\infty\frac{dk}{1-k^2}\ln 
k\cdot\ln^2\frac{\sqrt{k^2+4}+k}{2} 
=-{\pi^2}\cdot[\ln^2\frac{1+\sqrt5}{2}+\frac{\pi^2}{24}], 
\]

were used. These integrals were calculated in \cite{eks1}, and, to the best 
of our knowledge, do not appear in the mathematical handbooks.

Collecting all terms in the Table we obtain the analytic expression for the 
contribution to the Lamb shift induced by the gauge invariant set of 
diagrams in Fig.1c

\begin{equation}
\Delta 
E=\left(\frac{8}{3}\ln^2\frac{1+\sqrt5}{2}-\frac{872}{63}
\sqrt5\cdot\ln\frac{1+\sqrt5}{2}  +\frac{628}{63}\ln2-
\frac{2\pi^2}{9}+\frac{67282}{6615}\right) 
\end{equation}
\[
\times\frac{\alpha^2\left(Z\alpha\right)^5}{\pi 
n^3}\:{m}\left(\frac{m_r}{m}\right)^3.
\]

The numerical value of the factor in the parenthesis nicely coincides with 
the previous numerical results \cite{eg,pach1} within the accuracy of those 
numerical results. 

We would like to mention that exactly the same characteristic combinations 
of logarithms and square roots of five emerged earlier in the contribution 
of the same diagram to HFS. The result above differs from the respective 
contribution to HFS only by the values of the factors before these 
characteristic structures.

We are now able to give the state of the art expression for the total 
contribution to the Lamb shift of order $\alpha^2(Z\alpha)^5$, combining the
analytic results for the diagrams in Fig.1a-1b \cite{lap}, Fig.1c (this 
work), and the most precise numerical results for the other diagrams 
\cite{es,pach1}

\begin{equation}  \label{total}
\Delta E_{tot}=\left(\frac{8}{3}\ln^2\frac{1+\sqrt5}{2}-\frac{872}{63}
\sqrt5\cdot\ln\frac{1+\sqrt5}{2}  +\frac{680}{63}\ln2-
\frac{2\pi^2}{9}-\frac{25\pi}{63}\right.
\end{equation}
\[
\left.
+\frac{24901}{2205}
-7.920(1)\right) 
\frac{\alpha^2\left(Z\alpha\right)^5}{\pi 
n^3}\:{m}\left(\frac{m_r}{m}\right)^3
=-6.861(1)\;\frac{\alpha^2\left(Z\alpha\right)^5}{\pi 
n^3}\:{m}\left(\frac{m_r}{m}\right)^3.
\]

The phenomenological consequences of this result were discussed in detail, 
e.g., in \cite{es}, and we will not reproduce the discussion here. On the 
theoretical side we would like to mention that the analytic result above
nicely confirms qualitative arguments presented in \cite{es} about the 
natural scale of the order $\alpha^2(Z\alpha)^5$ contributions to the Lamb 
shift.

We are convinced that at the current state of the art the diagrams in 
Figs.1d-f do not admit analytic calculation for either the case of HFS 
or the case of the Lamb shift. In this sense we believe that the 
semianalytic result in Eq.(\ref{total}) cannot be improved further, by 
replacing numerical contributions by analytical results.

\bigskip
M. E. and V.S. are deeply grateful for the kind hospitality of the Physics 
Department at Penn State University, where this work was performed. The 
authors appreciate the support of this work by the National Science 
Foundation under grant number PHY-9421408.

\begin{table}
\caption{Individual Contributions to the Lamb Shift}
\begin{tabular}{lll}    
$\delta\epsilon_{11}$      &   $\frac{1}{64}$    & 
\\ 
$\delta\epsilon_{12}$      &   $-\frac{1}{256}$    & 
\\ 
$\delta\epsilon_{21}$      &   $-\frac{1}{8}\ln2-\frac{1}{32}$    & 
\\ 
$\delta\epsilon_{22}$      &   $\frac{1}{32}\ln2+\frac{3}{256}$    & 
\\ 
$\delta\epsilon_{31}$      &   $\frac{1}{4}\sqrt5\cdot\ln\frac{1+\sqrt5}{2}
-\frac{1}{4}\ln2-\frac{1}{16}$ & 
\\ 
$\delta\epsilon_{32}$      &   $-\frac{5}{12}\sqrt5\cdot\ln\frac{1+\sqrt5}{2}
+\frac{11}{24}\ln2+\frac{35}{288}$ & 
\\ 
$\delta\epsilon_{41}$      &   $\frac{7}{48}$    & 
\\ 
$\delta\epsilon_{42}$      &   $-\frac{113}{2880}$    & 
\\ 
$\delta\epsilon_{51}$      &   $-\frac{1}{96}$    & 
\\ 
$\delta\epsilon_{52}$      &   $\frac{13}{5760}$    & 
\\ 
$\delta\epsilon_{61}$      &   $-\frac{1}{160}$    & 
\\ 
$\delta\epsilon_{62}$      &   $\frac{37}{26880}$    & 
\\ 
$\delta\epsilon_{71}$      &   $\frac{1}{8}\ln^2\frac{1+\sqrt5}{2}
-\frac{1}{4}\sqrt5\cdot\ln\frac{1+\sqrt5}{2}
+\frac{1}{4}\ln2-\frac{\pi^2}{96}+\frac{1}{8}$ & 
\\ 
$\delta\epsilon_{72}$      &   $-\frac{1}{24}\ln^2\frac{1+\sqrt5}{2}
+\frac{2}{9}\sqrt5\cdot\ln\frac{1+\sqrt5}{2}
-\frac{17}{72}\ln2+\frac{\pi^2}{288}-\frac{19}{216}$ & 
\\ 
$\delta\epsilon_{81}$      &   $-\frac{5}{12}\sqrt5\cdot\ln\frac{1+\sqrt5}{2}
+\frac{5}{12}\ln2+\frac{5}{36}$ & 
\\ 
$\delta\epsilon_{82}$      &   $\frac{5}{12}\sqrt5\cdot\ln\frac{1+\sqrt5}{2}
-\frac{73}{160}\ln2-\frac{7361}{57600}$ & 
\\ 
$\delta\epsilon_{91}$      &   $\frac{5}{6}\sqrt5\cdot\ln\frac{1+\sqrt5}{2}
-\frac{59}{60}\ln2-\frac{2677}{14400}$ & 
\\ 
$\delta\epsilon_{92}$      &   $-\frac{15}{14}\sqrt5\cdot\ln\frac{1
+\sqrt5}{2}
+\frac{1013}{840}\ln2+\frac{219329}{705600}$ & 
\\ 
\tableline
$\delta\epsilon_{tot}$      &   $\frac{1}{12}\ln^2\frac{1+\sqrt5}{2}
-\frac{109}{252}\sqrt5\cdot\ln\frac{1+\sqrt5}{2}
+\frac{157}{504}\ln2-\frac{\pi^2}{144}+\frac{33641}{105840}$ & 
\\ 
\end{tabular} 
\label{table1} 
\end{table}

\begin{figure}        
\caption{Six gauge invariant sets of graphs producing corrections of order
$\alpha^2(Z\alpha)^5$.\label{fig1}}
\end{figure}

\end{document}